\documentclass[pre,aps,twocolumn,supergroupedaddress,epsfig,amsmath,amssymb,eqsecnum,floatfix]{revtex4}
\usepackage{epsfig,amsmath,amssymb,bm,epsf,graphics,psfrag,verbatim,framed}
\usepackage[usenames,dvipsnames]{color}
\usepackage{bbm}
\usepackage{mathrsfs}
\usepackage{amsfonts}
\usepackage{color}
\usepackage{{subfig}}
\usepackage{color}

\usepackage{epstopdf}

\begin{document}



\title[Strongly Coupled Titratable Macroions]{Titratable Macroions in Multivalent Electrolyte Solutions: 
Strong Coupling Dressed Ion Approach}

\author{Nata\v sa Ad\v zi\' c}
\email[]{natasa.adzic@ijs.si}

\affiliation{Department of Theoretical Physics, J. Stefan Institute, 1000 Ljubljana, Slovenia.}

\author{Rudolf Podgornik}
\affiliation{Department of Theoretical Physics, J. Stefan Institute, and
Department of Physics, Faculty of Mathematics and Physics, University of Ljubljana, 1000 Ljubljana, Slovenia.}

\date{\today}

\begin{abstract}
We present a theoretical description of the effect of polyvalent ions on the interaction between titratable  macro-ions. The model system consists of two point-like macro-ions with dissociable sites, immersed in an asymmetric ionic mixture of monovalent and polyvalent salts. We formulate a {\em dressed ion strong coupling theory}, based on the decomposition of the asymmetric ionic mixture into a weakly electrostatically coupled monovalent salt, and into polyvalent ions that are strongly electrostatically coupled to the titratable macro-ions. The charge of the macroions is not considered as fixed, but is allowed to respond to local bathing solution parameters  (electrostatic potential, $pH$ of the solution, salt concentration) through a simple {\em charge regulation} model. The approach presented, yielding an effective polyvalent-ion mediated interaction between charge-regulated macro-ions at various solution conditions, describes the strong coupling equivalent of the Kirkwood-Schumaker interaction.
\end{abstract}

\maketitle

\section{Introduction}

Charged colloidal particles such as proteins \cite{Warshel}, surfactant micelles and vesicles \cite{Micelles}, and nanoparticles \cite{Nanoint} are seldom describable as possessing a fixed charge or a fixed potential, though this notion does not cease to be popular~\cite{VO}. A more realistic point of view considers colloidal particles immersed in an aqueous electrolyte solution as possessing ionizable surface groups that respond to the local solution conditions~\cite{Borkovec1,Borkovec2}. Formally this perspective is equivalent to the assumption, that one can characterize the chargeable surface of the colloid particles with a specific free energy describing the dissociation/association equilibrium of surface ionizable groups or adsorption/desorption equilibrium of charged ions from solution to the surface \cite{CTAB}, and is referred to as {\em charge regulation} (CR). The first formalization of charge regulation was proposed in a seminal work by Ninham and Parsegian in the 1970's~\cite{NP-regulation} and formulated within the Poisson-Boltzmann (PB) theory of electrostatic interactions \cite{Safinya}. 

The implementation detais of the CR paradigm can vary. Chemical dissociation equilibrium of surface binding sites with the corresponding law of mass action was introduced already in ~\cite{NP-regulation} and was later generalizaed in different contexts~\cite{Regulation2,Regulation3,Regulation4,Regulation5}. A surface-site partition function or indeed a surface free energy model leads to the same basic self-consistent boundary conditions for surface dissociation equilibrium, but without an explicit connection with the law of mass action \cite{CTAB,Olvera, Olvera2,Natasa1,Natasa2,maggs,epl1,epl2,diamant, maarten,Ben-Yaakov1,Ben-Yaakov2}. The relationship between various boundary conditions that can be derived was elucidated recently \cite{epl2}.

Charge regulation has been invoked and widely applied in the context of various colloidal systems: stability and inter-surface forces due to the electrostatic double-layers \cite{stab,CTAB}, dissociation of amino acids and the corresponding electrostatic protein-protein interactions~\cite{Leckband1,Lund,BoJ,Fernando}, charge regulation of protein aggregates and viral shells \cite{Nap}, and of polyelectrolytes and polyelectrolyte brushes~\cite{Netz-CR,Borukhov,Kilbey,Zhulina}, as well as charge regulation of charged lipid membranes~\cite{membranes1,membranes2,membranes3}. Here, we specifically dedicate ourselves to the problem of the connection between charge regulation and electrostatic interactions between proteins in ionic solutions \cite{Warshel,Leckband2}. We recently showed how the Kirkwood-Schumaker (KS) interaction \cite{KS1,KS2} follows directly from charge regulation, based on different surface free energy models~\cite{Natasa1,Natasa2,maggs}, and presented a theory of fluctuation interaction between macroions subject to charge regulation, thereby generalizing the KS perturbation approach \cite{KS1,KS2}. 

We formulated this generalized KS problem by decoupling the system composed of two charge-regulated macroions and an intervening bathing ionic solution into two parts: the solution part and the surface part ~\cite{Natasa1,Natasa2,maggs}. These were then treated within separate approximation schemes. The solution part was treated on the linearized weak-coupling Debye-H\" uckel (DH) level \cite{RudiandCo}, while the surface part was shown to be amenable to an exact evaluation. This decomposition allowed us to derive a closed-form expression for the total effective interaction between macroions that we were able to connect with the original KS expression. In fact, our generalized fluctuation-mediated interaction reduces exactly to the KS result in the limit of large separations between macroioins and in fact presents a one-loop (Gaussian fluctuation) correction to the mean-field DH result. As such, it is only valid for a weakly charged system, where the salt ions mediating the mean-field as well as fluctuation interactions are electrostatically weakly coupled to the macroions. No such approximations were necessary in a 1D model that can be evaluated exactly \cite{maggs} and supports the conclusions based on the weak coupling (WC) approximation.

We now change the perspective and consider a case where the bathing solution contains not only weakly charged monovalent salt ions, but also polyvalent ions that are strongly electrostatically coupled  to the charged macroions, mediating the interaction between them. A possible realization would correspond to a mixture of multivalent ions in a bathing solution of monovalent ions, a situation rather typical in the context of e.g. semiflexible biopolymers, where multivalent ions are believed to play a key role in their condensation \cite{dressed1,dressed2}. With the presence of polyvalent ions in the system, the WC paradigm in general breaks down and the existence of KS interactions becomes dubious~\cite{RudiandCo}. However, there exists a theory, the {\em dressed ion theory}, based on an asymmetric treatment of the different components of the bathing electrolyte solution, that would allow us to analyze the effect of charge reguation of macroions also in the presence of polyvalent salt ions in the bathing solution~\cite{dressed1,dressed2}. It is based on the fact, that one can use the WC DH approach in order to describe the monovalent salt ions, while a strong coupling (SC) approach is preferable for the polyvalent ion part. This combined {\em weak-strong coupling} approach~\cite{dressed1,dressed2} effectively leads to dressed interactions between polyvalent ions and thus also affects the interactions mediated by polyvalent counterions between two like charge-regulated macroions. The ensuing  effective interactions between macroions would then correspond to a {\em generalized KS interaction}, mediated by strongly coupled salt ions and not by weakly coupled monovalent salt. This generalized KS interaction would consequently also cease to be fluctuational in nature, i.e. of the type proposed in the original work of Kirkwood and Schumaker \cite{KS1,KS2}, but would show a different behavior stemming from the polyvalent ion mediated interactions coupled to the charge regulation response of the dissociation equilibrium at the macroion surfaces.

Our approach as detailed below, is composed of disjoined parts brought together to describe this new type of generalized KS interaction,  and a short {\em guided tour} through the conceptual and calculational flowchart is thus in order. The dissociable surfaces of the two identical macroions, representing two proteins with dissociable aminoacids, are described with a charge-regulation surface free energy that allows the effective charge to vary between a positive and a negative maximal value. We then contract the macroion to a point particle merely as a calculational device, since we can then disregard the angular distribution of the dissociable groups along the surface, remaining solely with the monopolar charge as the only characteristics of the macroion. The bathing solution for the macroions, assuming to be an ionic mixture of monovalent salt and polyvalent ions, is then treated within the {\em dressed ion theory}, i.e. the monovalent salt is described within the WC and the polyvalent ions within the SC paradigm, an approximate approach that has already proved valuable in other contexts~\cite{dressed1,dressed2}. We then further approximate the non-linear surface charge regulation free energy with a Gaussian expansion proved to be a good description on the WC level ~\cite{Natasa1,Natasa2}.  Finally, we study the obtained expressions for the effective generalized KS interaction between the macroions in the various parts of the parameter space and comment on the results. 

The dressed ion theory, as a variant of the SC theory~\cite{RudiandCo}, does not hold the same status as the {\em original SC theory}, valid {\em exactly} for a counterion-only system in the limit of large coupling constant \cite{Netz01,AndreEPJE, hoda_review}. In fact the regime of validity of this approach can be only checked against explicit-ion Monte Carlo simulations, showing that the dressed ion theory  can indeed give quantitatively accurate results in a wide range of realistic parameter values \cite{dressed1,dressed2,SCdressed3}. 

\section{General formalism}

\subsection{Model}

\begin{figure}
\centering{\includegraphics[width=0.48\textwidth]{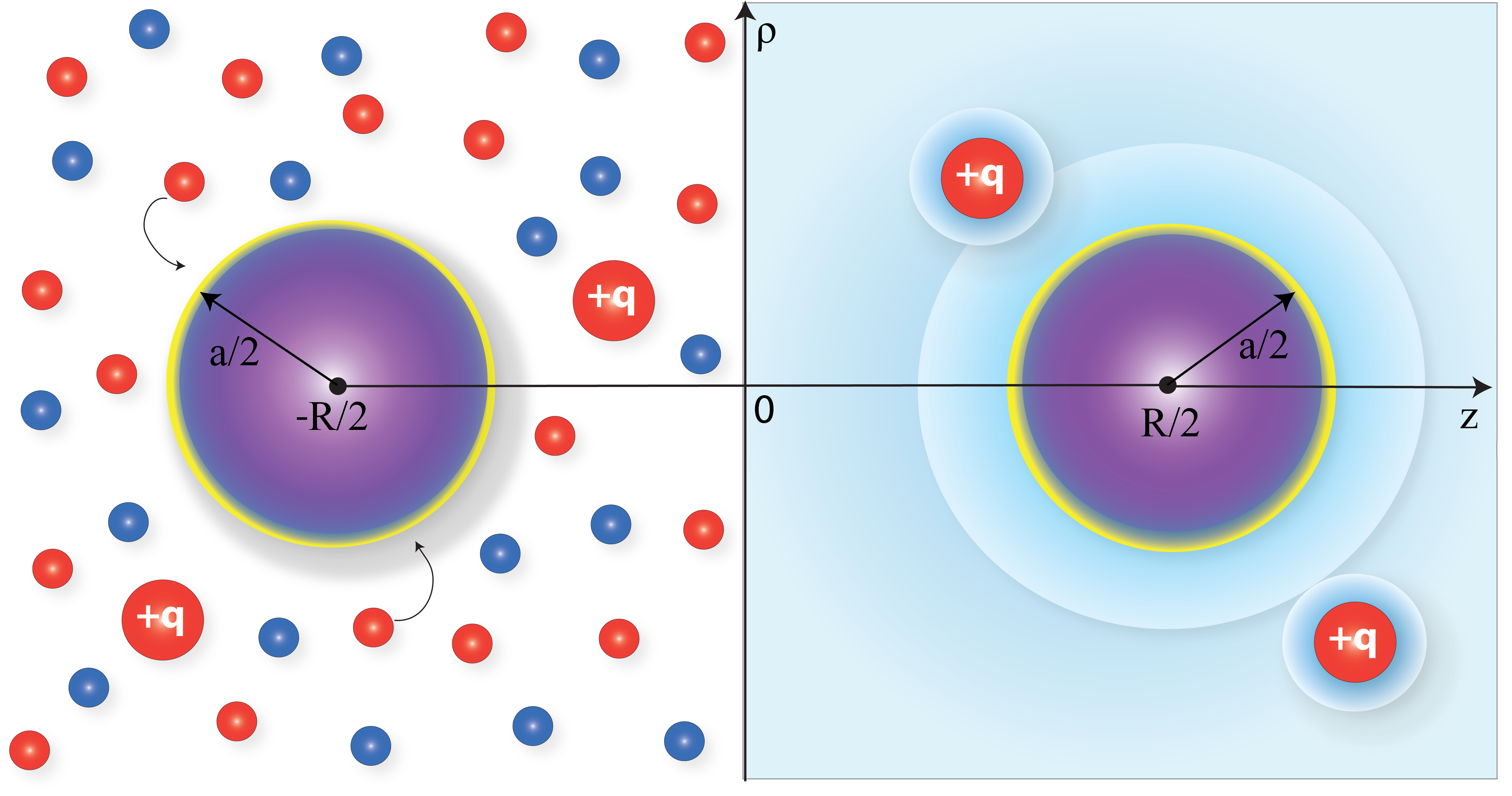}}
\caption{Shematic representation of the model: two charge regulated macro-ions, representing two proteins with titratable surface groups,  immersed in a mixture of monovalent-polyvalent salt solution. The microscopic model (left) shows the different types of ions and the surface dissociation equilibrium on the surface of the macroion. The coarse-grained dressed ion model (right) shows the effective DH potential (light coloured corona) of the macroioin as well as the polyvalent solution ions.  In a cylindrical  coordinate system with the $z$-axis connecting the two macroions, having its origin in the middle between the macroions,  the macroions are located at ${\bf r}_1 = (x, y, -R/2)$, and  ${\bf r}_2 = (x, y, R/2)$, respectively. }
\label{fig:fig1}
\end{figure}

The system under consideration consists of two equal titratable macroions immersed in a bathing solution, itself composed of a mixture of monovalent salt ions as well as polyvalent ions of valency $q$, see Fig. \ref{fig:fig1}. Two macroions, representing two titratable proteins, are located at ${\bf r}_1$ and ${\bf r}_2$ so that their separation is equal to $\vert {\bf r}_1 - {\bf r}_2 \vert = R$. The macroions are assumed to be identical with a radius of $a/2$ and can have either sign. Furthermore, the macroions are charge-regulated with adsorption sites which can exchange a proton from the environment, and are described with the lattice gas free energy, see below, with a site number coefficient of $\alpha = 2$. Thi simplies that there are twice as many proton adsorption/dissociation sites as there are negative fixed charges. This allows the total charge of the macroion to span negative as well as positive values, a basic tenet of our charge regulation model. 

The macroion charge is thus not fixed, but responds to the local solution conditions. We also assume that the macroions are "small" in the specific sense that the angular variation of the local electrostatic potential along their surface is negligible. This implies that we only deal with effective monopolar fluctuations, disregarding the subdominant higher multipolar fluctuations that would correspond to a generalization of the full van der Waals interaction potential \cite{French}. The higher multipolar KS interactions remain as a possible future topic of our investigation.

\subsection{Charge regulation}

For charge-regulated titratable macroions we have recently introduced several models~\cite{Natasa1,Natasa2}, based on a charge dissociation free energy that generalizes the law of mass action charge-regulation approach of Ninham and Parsegian ~\cite{NP-regulation}. In these models the charge regulation is described by a surface free energy $f_S({\bf r}) = f_S(\phi({\bf r}))$ that depends on the surface electrostatic potential $\phi({\bf r})$. For each macroion the total charge regulation free energy $F[\phi({\bf r})]$ would thus be a functional of the surface potential amount to
\begin{equation}
F[\phi({\bf r})] = \oint_{S} f_S(\phi({\bf r})) d^2{{\bf r}},
\label{cagh}
\end{equation}
where $S$ is the surface area of the macroion. At this point we simplify matters by furthermore assuming that the macroions are spherical and of vanishing radius, i.e. they are point particles. Of course this approximation will only work for sufficiently large separations between them and small separation regime would need to be analyzed separately. It will soon become clear why this type of approximation simplifies the calculation substantially.

Assuming then that the macroion is located at $({\bf r}_0)$ and has a vanishing radius $a \longrightarrow 0$, the integral of the dissociation free energy over the surface of the macroion, Eq. \ref{cagh}, simply gives a total dissociation energy of the point-like macroion as a {\sl function} of the local potential at the point ${\bf r} = {\bf r}_0$. The point-like approximation for the macroion therefore disregards the angular variation of the local electrostatic potential along the surface of the macroions and can describe only monopolar charge regulation, while higher multipoles are ignored.

In the next step one needs to assume a model for $f_S(\phi({\bf r}))$. We already invoked several models ~\cite{Natasa1,Natasa2} related to the original Ninham-Parsegian model~\cite{NP-regulation}. Focusing on a simple two-parameter model we introduce the following {\sl Ansatz} for a charge regulated point-like macroion~\cite{Natasa2}
\begin{eqnarray}
&&F(\phi({\bf r}_0)) = \lim_{a \rightarrow 0} \oint_S f_S(\phi({\bf r})) d^2{{\bf r}} \longrightarrow  \nonumber\\
&& - N e_0 \phi({\bf r}_0) - \alpha N k_BT \log{\left( 1 + b e^{-\beta e_0 \phi({\bf r}_0)}\right)},
\label{bdfhjwk}
\end{eqnarray}
where $\phi({\bf r}_0)$ is now the local electrostatic potential at the position of the ion, while $N$ and $\alpha$ are two parameters characterizing the dissociation process. The site number coefficient $\alpha$ quantifies the number of dissociation sites, and $\log{ b} = \beta \mu_S$, incorporates the free energy of charge dissociation $ \mu_S$. 

In the case of protonation of the titratable surface charge, it furthermore follows that $\log{ b} =  \log{10}(pH - pK)$, where $pK$ is the dissociation constant and $pH = -\log{[H^{+}]}$ is the proton concentration in the bulk, differing from the local value of $pH$ at the dissociation site ~\cite{Natasa1,Natasa2}. It is straightforward to see that the free energy Eq. \ref{bdfhjwk} is composed of the electrostatic energy of $N$ fixed negatively charged sites with the total charge $- N e_0$ and $\alpha N$ lattice gas sites, that can be filled with adsorbing protons from the solution; in fact $1 + e^{\mu}$ is nothing but the lattice gas partition function for single occupation sites, with zero energy for the empty site and $\mu$ for the filled site, while $\log{\left(1 + e^{\mu} \right)}$ is just the corresponding grand canonical surface pressure. 

The form of the charge regulation free energy then allows us to derive the effective charge of the charge-regulated macroion as a function of the local electrostatic potential in the form
\begin{equation}
e\left(\phi \right) = \frac{\partial F(\phi)}{\partial \phi}
\end{equation}
where $F(\phi)$ is the dissociation free energy Eq. \ref{bdfhjwk} yielding
\begin{equation}
e\left(\phi \right) = e_0 N\!\!\left(\left(\frac{\alpha }{2}-1\right)\!-\!\frac{\alpha }{2}\tanh{[-{\textstyle\frac{1}{2}}(\ln{b}-\beta e_0\phi )]} \right).
\label{vaeht}
\end{equation}
The effective charge of the macroion then varies in the interval $-Ne_0< e(\phi) < (\alpha - 1)Ne_0$. Choosing the site number coefficient to be $\alpha = 2$, one thus remains with a symmetric charge regulated macroion whose effective charge varies within the interval $-Ne_0 < e(\phi) < +Ne_0$. This is the generic charge regulation model that we will consider as a simple description of the protein charge regulation in what follows.

\subsection{Field Theory-general formalism}

We proceed by writing the partition function through the Hubbard-Stratonovich transform for the Coulomb potential as explained in detail elsewhere \cite{Rudifunint}. This leads to a field theory, where the classical partition function is represented as a functional integral over the fluctuating electrostatic potential. Two explicit exact limiting results are then obtainable from this representation in the case of a counterion-only system \cite{RudiandCo}: the saddle-point of this field theory in fact corresponds to the mean-field Poisson-Boltzmann (PB) approximation, while the Gaussian fluctuation correction together with the  PB theory constitutes the {\sl WC theory}; the first order virial expansion of the partition function then constitutes the {\sl SC theory}, unrelated to the PB approximation. The latter can be further generalized in the case of a mixed system by treating the monovalent salt on the WC level while the polyvalent ions are described on the SC level, i.e. their contribution to the partition function is written as a second order virial expansion theory. This approximation was dubbed the "dressed ion theory" ~\cite{dressed1,dressed2}.

Assuming that the fluctuating electrostatic potential of the macroions is $\phi ({\bf r} = {\bf r}_1) = \varphi _1$ and of the other one is $\phi ({\bf r} = {\bf r}_2) = \varphi _2$, located at $\ {\bf r}_1$ and  $\ {\bf r}_2$ respectively, the partition function of the system within the dressed ion theory can be derived in the field-theoretic form as \cite{Natasa1,Natasa2}
\begin{equation}
{\cal{Z}}=
\int\!\!\int d\varphi _1e^{-\beta F(\imath \varphi _1)} G(\varphi_1,\varphi_2) e^{-\beta F(\imath \varphi _2)}d\varphi _2,\label{eq:path}
\end{equation}
where $F(\imath \varphi)$ is charge regulation free energy, Eq. \ref{bdfhjwk}, evaluated at imaginary values of the fluctuating electrostatic potential, and the field propagator or the Green function, giving the probabilty of field configurations with $\phi ({\bf r} = {\bf r}_1) = \varphi _1$ and $\phi ({\bf r} = {\bf r}_2) = \varphi _2$, is given by
\begin{equation}
G(\varphi_1,\varphi_2)=\int{\cal D}[\varphi({\bf r})]e^{-\beta H[\varphi ]}\delta(\varphi({\bf r}_1)-\varphi_1)\delta(\varphi({\bf r}_2)-\varphi_2),
\label{lnrs}
\end{equation}
with the bulk field action:
\begin{eqnarray}
-\beta H[\varphi]=-\beta H_0[\varphi] +  \!\lambda_c\!\!\int d{{\bf r}} e^{i\beta qe\varphi ({\bf r})}, 
\end{eqnarray}
where $\lambda_c$ is the fugacity of the polyvalent ions with valency $q$ and $H_0[\varphi] $ is the DH field Hamiltonian 
\begin{eqnarray}
-\beta H_0[\varphi]&=& {\textstyle\frac12} \epsilon\epsilon_0 \int d{\bf r}d{\bf r}' \varphi({\bf r}) u_{DH}^{-1}({\bf r}, {\bf r}' )\varphi ({\bf r} ' ) \nonumber\\
&&= {\textstyle\frac12} \epsilon\epsilon_0 \int ((\nabla \varphi ({\bf r}))^2+\kappa^2 \varphi^2({\bf r})) d{\bf r}.
\end{eqnarray}
Here we have assumed that the monovalent salt is weakly coupled to the rest of the charges and can be treated on the DH level. The inverse square of Debye length was introduced as $\kappa^2 = 4\pi \ell_{\mathrm{B}} n_\textrm b$, with $\ell_B$ the Bjerrum length and  $n_\textrm  b=2n_0+qc_0$, where $n_0$ is the bulk concentration of the monovalent salt and $c_0$ is the bulk concentration of the multivalent ions, assumed to originate in dissociation of a $q$:1 salt. The DH interaction kernel $u_{DH}^{-1}({\bf r}, {\bf r}' )$ implies a screened effective DH interaction potential
\begin{equation}
u_{DH}({\bf r}, {\bf r}' ) = \frac{1}{4\pi\epsilon\epsilon_0} \frac{e^{-\kappa \vert {\bf r}- {\bf r}' \vert}}{\vert {\bf r}- {\bf r}' \vert} = \frac{1}{4\pi\epsilon\epsilon_0} {\tilde u}_{DH}({\bf r}, {\bf r}' )
\end{equation}
between the polyvalent ions and the macroions. On this level the polyvalent ions are thus treated explicitly, but their interactions with the macroions are described with a dressed electrolyte-mediated effective DH potential. 

The strong asymmetry in the system, implied by the presence of polyvalent mobile ions, together with their small concentration leads straightforwardly to the virial expansion for their contribution to the partition function that yields to the lowest order~\cite{dressed1,dressed2,SCdressed3}
\begin{eqnarray}
e^{-\beta H[\varphi]}&=&e^{-\frac{1}{2}\beta \int d{\bf r}d{\bf r}' \varphi({\bf r}) u_{DH}({\bf r}, {\bf r}' )\varphi ({\bf r} ' )  }(1+\nonumber\\
&& \lambda_c\int_V d{\bf r}_0 e^{i\beta qe\varphi ({\bf r}_0)}+...),
\end{eqnarray} 
furthermore implying that the propagator $G(\varphi_1,\varphi_2)$ can be decomposed into
\begin{equation}
G(\varphi_1, \varphi_2)=G_0(\varphi_1, \varphi_2)+\lambda_c \int_V d{\bf r}_0 G_1(\varphi_1, \varphi_2; {\bf r}_0).
\end{equation}
The propagator $G_1(\varphi_1, \varphi_2; {\bf r}_0)$, describes the field propagation from macro-ion at ${\bf r}_1$ to macro-ion at ${\bf r}_2$ mediated by the presence of the polyvalent ion $q$ at ${\bf r}_0$ integrated over the fluctuating potential at the positions of both macroions. Formally this can be expressed as
\begin{widetext}
\begin{eqnarray}
G_1(\varphi_1,\varphi_2; {\bf r}_0)= \int{\cal D}[\varphi({\bf r})]\delta(\varphi({\bf r}_1)-\varphi_1)e^{-\beta H_1[\varphi; {\bf r}_0]}\delta(\varphi({\bf r}_2)-\varphi_2),
\label{ngcrlks}
\end{eqnarray}
\end{widetext}
where the effective field action $H_1[\varphi; {\bf r}_0] $ can be decomposed into the DH part due to the weakly coupled monovalent salt ions and the coupling between fluctuating potential and the polyvalent ion of valency $q$ located at ${\bf r}_0$, i.e.
\begin{equation}
\beta H_1[\varphi; {\bf r}_0] = \beta H_0[\varphi] - i\beta \int \rho({\bf r}_0) \varphi({\bf r}) d{\bf r}.
\end{equation}
The last term describes the interaction with the polyvalent ion with density $$\rho({\bf r}_0)=q\delta ({\bf r}-{\bf r}_0).$$This formal expression for the propagator $G_1(\varphi_1,\varphi_2; {\bf r}_0)$ is thus identical to the partition function of two macroions at positions ${\bf r}_{1,2}$ with set values of the fluctuating potential $\varphi_{1,2}$ interacting via the DH interaction with an additional point particle of charge $qe_0$ at ${\bf r}_0$ at the positions of the two point-like macroions. The functional integral in Eq. \ref{ngcrlks} simply indicates the summation over all fluctuating potentials that satisfy these constraints.

With these definitions the full dressed ion partition function can then be cast into the sum of two disjoint terms, one corresponding to two isolated polyvalent ions interacting directly via DH potential, and the other describing the polyvalent ion mediated interaction 
\begin{widetext}
\begin{eqnarray}
{\cal{Z}} = \int \int d\varphi _1e^{-\beta F(\varphi _1)}\Big[G_0(\varphi_1, \varphi_2)+ \lambda_c\int_V d{\bf r}_0  G_1(\varphi_1, \varphi_2; {\bf r}_0) \Big]e^{-\beta F{(\varphi _2)}}d\varphi _2={\cal{Z}}_0+\lambda _c{\cal{Z}}_1\nonumber\\
~
\label{eq:path}
\end{eqnarray}
\end{widetext}
with obvious definitions for the two terms in the sum. ${\cal{Z}}_0$ and ${\cal{Z}}_1$ by definition then give the zero order and the first order polyvalent ion virial expansion contributions in the partition function. ${\cal{Z}}_0$ has been already analyzed in \cite{Natasa2} and ${\cal{Z}}_1$ will be evaluated below. The above decomposition of the full partition function ${\cal{Z}} = {\cal{Z}} (R)$ is the essence of the dressed ion theory and the corresponding free energy will describe the interactions between the two macroions as a function of their separation and model parameters. 

\subsection{Dressed ion theory and charge regulation}

The first order virial expanded Green function $G_1(\varphi_1, \varphi_2; {\bf r}_0)$ can be reduced to Gaussian functional integrals, see Appendix \ref{subsec:app1}, and can be derived in an explicit form
\begin{eqnarray}
&&G_1(\varphi_1,\varphi_2; {\bf r}_0)=\frac{\exp{\left( \frac{1}{2} \tilde\Phi_i({\bf r}_0)
{G^{-1}}_{ij}({\bf r}_1,{\bf r}_2) \tilde\Phi_j({\bf r}_0)\right)}}{\sqrt{\det{G_{ij}({\bf r}_1,{\bf r}_2)}}} \nonumber\\
~\label{eq:prp1}
\end{eqnarray}
where we introduced
\begin{equation}
\Phi_i({\bf r}_0) = i\varphi_1 + qe_0 u_{DH}({\bf r}_0,{\bf r}_i), 
\end{equation}
and 
\begin{equation}
{G}_{ij}({\bf r}_1,{\bf r}_2) = {k_BT} \Bigg(\begin{matrix}
{a}^{-1} & u_{DH}({\bf r}_1,{\bf r}_2) \\
u_{DH}({\bf r}_1,{\bf r}_2) & {a}^{-1}
\end{matrix}\Bigg),
\label{eq:g1}
\end{equation}
for $i,j =1,2$. From the above expressions it is clear that the macroions interact with themselves as well as with the polyvalent ion whose position within the system will be finally integrated over.    The terms with ${a}^{-1}$ describe the self-interaction of the macroions with diameter $a$, while the interaction between the macroions as well as between the macroions and the polyvalent ion are given by the DH screened interaction potential. In a cylindrical  coordinate system with the $z$-axis connecting the two macroions, having its origin in the middle between the macroions, themselves separated by $R$, the position of the polyvalent ion with respect to both macroions can be written as $\vert {\bf r}_0 - {\bf r}_1 \vert=\sqrt{\rho_0^2+(R/2+z_0)^2}$, and  $\vert {\bf r}_0 - {\bf r}_2 \vert=\sqrt{\rho_0^2+(R/2-z_0)^2}$, respectively. 

Going back to the definition of the partition function ${\cal{Z}}_1$, Eq. \ref{eq:path}, one can finally write
\begin{equation}
{\cal{Z}}_1=\!\! \lambda_c\int_V\!\! d{\bf r}_0 \int \!\!d\varphi _1 e^{-\beta F(\imath\varphi _1)}\Big[ G_1(\varphi_1, \varphi_2) \Big]e^{-\beta F{(\imath \varphi _2)}} d\varphi _2.
\label{eq:pathz1}
\end{equation}
While the Green function $G_1(\varphi_1, \varphi_2) $ is Gaussian in the two fields, the surface field action $F(\imath\varphi )$ is not. Additional considerations are therefore needed to proceed. First we note, as amply elucidated in Ref. \cite{Natasa2}, that an {\sl exact} method of evaluation of ${\cal{Z}}_1$ is available if one expands the surface field action into a series, yileding 
\begin{eqnarray}
e^{-\beta F(\imath\varphi ) }&=&e^{-i \beta N e_0\varphi }(1+b e^{i\beta e_0\varphi})^{2N}=\nonumber\\
&&\sum_{n=0}^{2N}\left(\begin{array}{c} 2N \\ n \end{array}\right)e^{-i \beta N e_0\varphi }b^{n}e^{i\beta e_0n\varphi}.
\label{bucek}
\end{eqnarray}
While the above expansion, giving a sum over surface terms linear in the fluctuating potential, could in principle be used for a direct numerical evaluation of the partition function, we have already shown \cite{Natasa1,Natasa2} that an additional approximation, simplifying the calculation extensively, yields an accurate result that compares well with the exact summation. This further step relies on the Gaussian approximation for the binomial coefficient in the above expansion
\begin{equation}
\lim_{N \gg 1}\left(\begin{array}{c} 2N \\ n \end{array}\right) \simeq \frac{2^{2 N}}{\sqrt{\pi N}}e^{-\frac{( N-n)^2}{ N}},
\end{equation}
valid strictly in the limit of a large number of adsorption sites, $N \gg 1$. Introducing the auxiliary fields $x_1= N-n_1$ and $x_2= N-n_2$, summation in Eq. \ref{bucek} can thus be replaced with an integration, so that the partition function assumes a much simplified and easily calculable form
\begin{widetext}
\begin{eqnarray}
{\cal{Z}}_1=\!\! \lambda_c   \int_V\!\! d{\bf r}_0 \int \!\!d\varphi _1 d\varphi _2 \int \!\! dx_1 dx_2   \frac{e^{s(q^2;{\bf r}_1,{\bf r}_2) + \ln{10}\ (pH-pK)(x_1+x_2) -\frac{x_1^2}{N}-\frac{x_2^2}{N}}}{\sqrt{\det{{G}_{ij}({\bf r}_1,{\bf r}_2)}}}  \exp{\!\Bigg[\!\!-\frac{1}{2} \varphi_i  {G^{-1}}_{ij}({\bf r}_1,{\bf r}_2) \varphi_j  + \imath \beta e_0\varphi_i (-x_i+q y_i) \Bigg]}. \nonumber\\
~
\label{eq:pathz12}
\end{eqnarray}
\end{widetext}
Here the effective interaction matrix $ {G}_{ij}({\bf r}_1,{\bf r}_2)$ has been already defined in Eq. \ref{eq:g1}, while ${G^{-1}}_{ij}({\bf r}_1,{\bf r}_2)$ is its matrix inverse. In addition we introduced two additional auxiliary fields with no other role but to make the notation more compact,
\begin{equation}
y_1=\frac{a^2}{1-\frac{a^2}{R^2}e^{-2\kappa R}}\left(\frac{1}{a}{\tilde u}_{DH}({\bf r}_0,{\bf r}_1)-\frac{e^{-\kappa R}}{R}{\tilde u}_{DH}({\bf r}_0,{\bf r}_2)\right)
\end{equation}
and
\begin{equation}
y_2=\frac{a^2}{1-\frac{a^2}{R^2}e^{-2\kappa R}}\left(\frac{e^{-\kappa R}}{R}{\tilde u}_{DH}({\bf r}_0,{\bf r}_1)-\frac{1}{a}{\tilde u}_{DH}({\bf r}_0,{\bf r}_2) \right),
\end{equation}
The effective self-energy of the polyvalent ion, $s(q^2;{\bf r}_1,{\bf r}_2)$, mediated by both macroions, is proportional to the square of the polyvalent ion charge and is given by
\begin{widetext}
\begin{eqnarray}
s(q^2; {\bf r}_1,{\bf r}_2)=\frac{1}{2}q^2 \frac{l_B a }{1-\frac{a^2}{R^2}e^{-2\kappa R}}\times \Big({\tilde u^2}_{DH}({\bf r}_0,{\bf r}_1)+{\tilde u^2}_{DH}({\bf r}_0,{\bf r}_2) -2\frac{a}{R}e^{-\kappa R}{\tilde u}_{DH}({\bf r}_0,{\bf r}_1){\tilde u}_{DH}({\bf r}_0,{\bf r}_2) \Big).\nonumber\\
~
\end{eqnarray}
\end{widetext}
After integrating out the $x_i$-auxiliary fields and the fluctuating potentials of the two macroions, $\varphi _1, \varphi _2$, one obtains the final result in the form of an integration over the position of the polyvalent ion
\begin{widetext}
\begin{eqnarray}
{\cal{Z}}_1 &=& {\cal{Z}}_0 \lambda_c  \int_V\!\! d{\bf r}_0 \times \exp{\!\!\Bigg[\!\frac{q \ell_B N(pH\!-\!pK)\ln{10} \left( {\tilde u}_{DH}({\bf r}_0,{\bf r}_1)\! +\!{\tilde u}_{DH}({\bf r}_0,{\bf r}_2) \right) }{2+Nl_B\frac{1}{a} [1+\frac{a}{R}e^{-\kappa R}]}\!\!\Bigg]} \nonumber\\
&& \times \exp{\Bigg[\frac{1}{2}q^2 \ell^2_B \Big(C_{11}{\tilde u}_{DH}^2({\bf r}_0,{\bf r}_1) + C_{22} {\tilde u}_{DH}^2({\bf r}_0,{\bf r}_2)-} {2C_{12} {\tilde u}_{DH}({\bf r}_0,{\bf r}_1){\tilde u}_{DH}({\bf r}_0,{\bf r}_2)  \Big) \Bigg]},\nonumber\\
~\label{eq:pf1}
\end{eqnarray}
\end{widetext}
where ${\cal{Z}}_0$ is the partition function of a system of two isolated charge-reguated macroions on the WC approximation level, already derived within the context of the weakly coupled macroions in monovalent salt solution \cite{Natasa1,Natasa2} and given by
\begin{equation}
{\cal{Z}}_0 = \frac{\exp{\Big[\frac{N[ (pH-pK)\ln{10}]^2}{2+Nl_B\frac{1}{a}[1+\frac{a}{R}e^{-\kappa R}]}\Big]}}{\sqrt{\frac{4}{N^2}+\frac{2}{N}\frac{l_B}{a}+\frac{l_B^2}{a^2}[1-\frac{a^2}{R^2}e^{-2\kappa R}]}}. 
\label{aeknux}
\end{equation}
Above we also introduced the generalized self and mutual capacitances as
\begin{eqnarray}
C_{11}&=&C_{22}=\frac{\frac{l_B }{a}+\frac{2}{N} }{\left( \frac{l_B}{a}+\frac{2}{N}  \right)^2-\frac{e^{-2\kappa R}}{R^2/l_B^2}} ;\nonumber\\
C_{12}&=&\frac{\frac{l_B e^{-\kappa R}}{R}}{\left( \frac{l_B}{a}+\frac{2}{N}  \right)^2-\frac{e^{-2\kappa R}}{R^2/l_B^2}}.\label{eq:cap}
\end{eqnarray}
While they do not have the standard form of the capacitances, since they both contain also contributions from mutual interactions, in the limit of large separations between the macroions they do reduce to the expected values. The difference in the definition of capacitances is a consequence of the fact that the dressed ion theory is not Gaussian as far as the fluctuating potential is concerned, in contrast to the WC case analyzed before \cite{Natasa1,Natasa2}, but is a non-linear SC theory. Capacitance is a WC concept, pertaining to Gaussian fluctuations and thus does not have a direct equivalent in the SC theory.

\begin{figure*}
\centering{\subfloat[]{\includegraphics[width=0.34\textwidth]{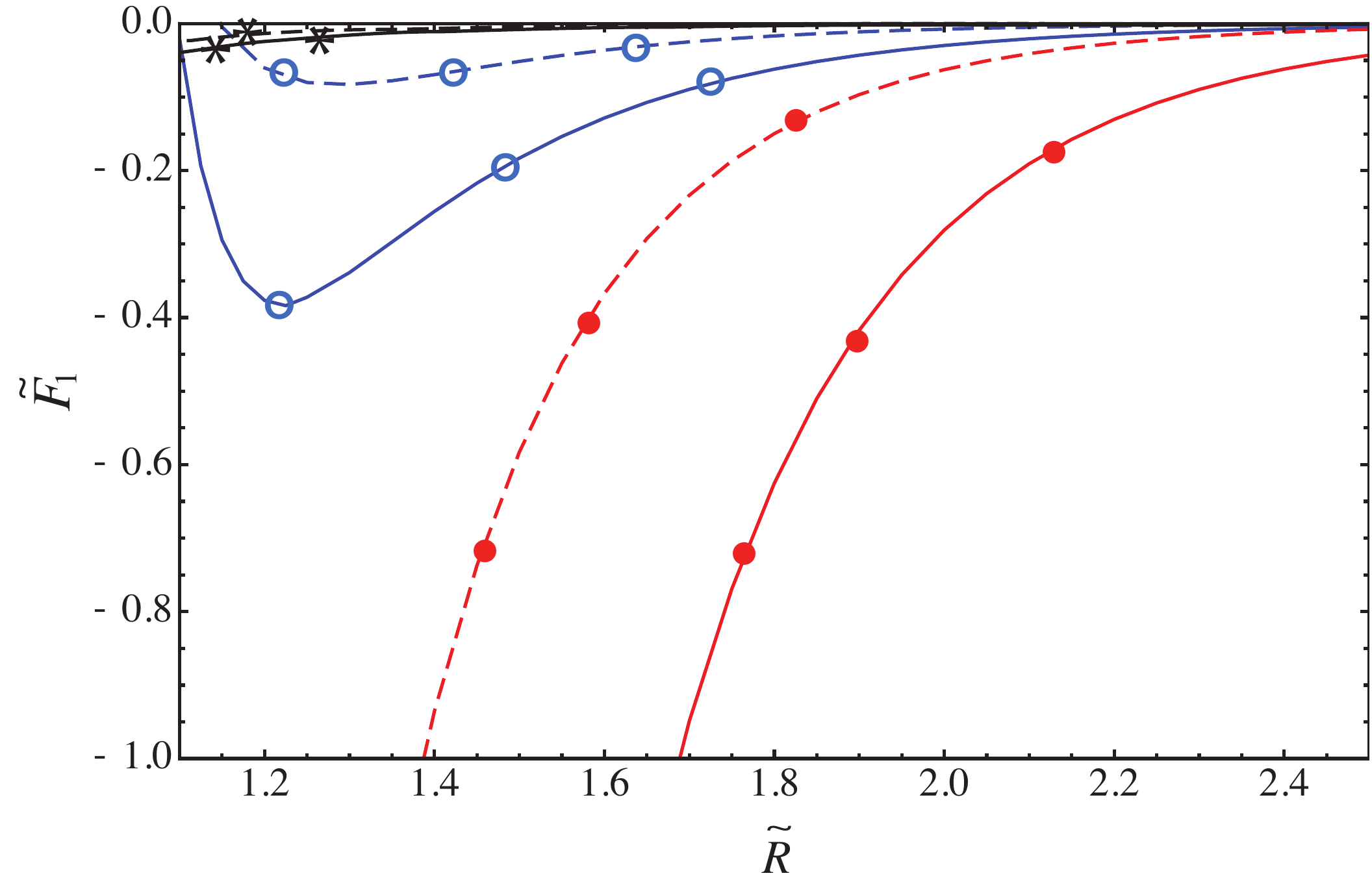}}\subfloat[]{\includegraphics[width=0.33\textwidth]{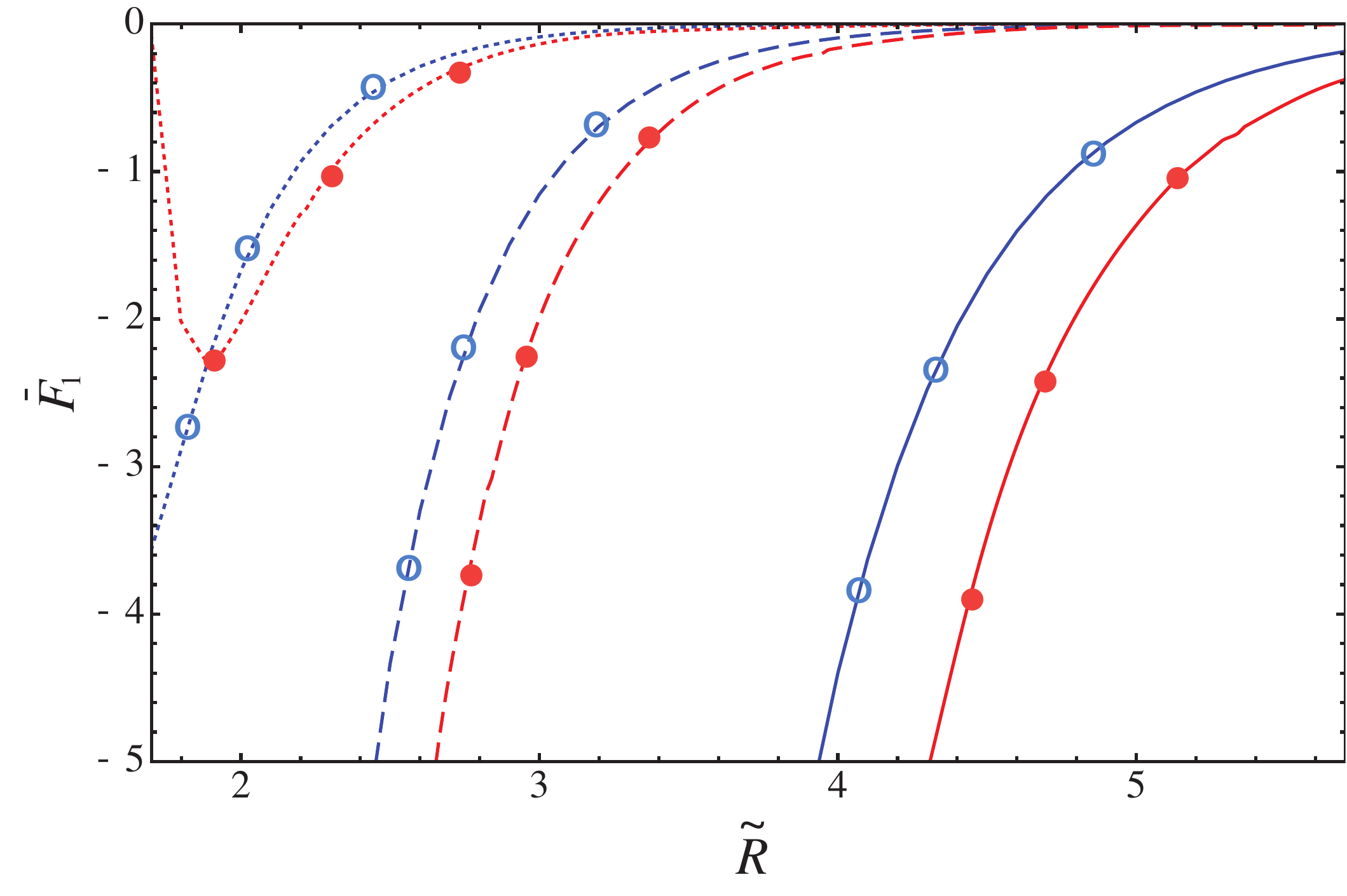}}\subfloat[]{\includegraphics[width=0.34\textwidth]{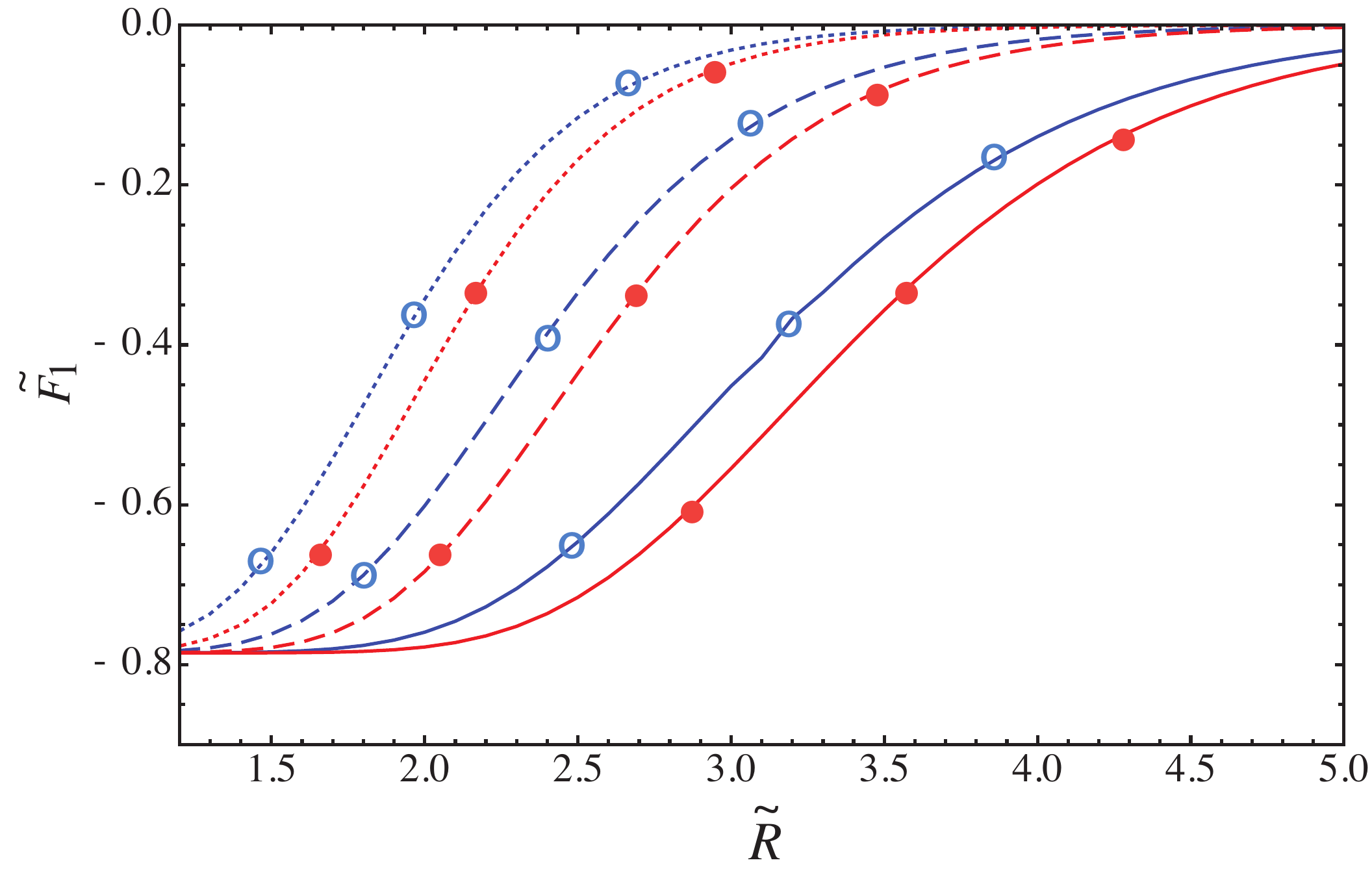}}
\caption{Interaction free energy contribution ${\cal{\tilde{F}}}_1(R)$, Eq. \ref{eq:f1}, originating in the presence of polyvalent ions, for (a) $pH-pK=0$; (b)  $pH-pK=3$ and (c) $pH-pK=-3$. Blue lines (marked with an open circle) $q=3$, red lines (marked with a filled circle) $q=4$, black lines (marked with star) $q=0$ (standing for the attraction coming from Eq. \ref{eq:ks0}). Solid lines correspond $n_0=150$ mM, dashed $n_0=300$ mM, while dotted stand for  $n_0=500$ mM in (b) and (c), while in (a) monovalent salt concentration is chosen as $n_0=100$ mM solid lines and $n_0=150$ mM dashed lines. Macroions diameter $a=1$ nm, number of adsorption sites $N=7$ and $c_0=1$ mM.}\label{fig:fig2}}
\end{figure*}

We now write down the free energy difference between the state where the two macroions are at a finite spacing $R$ and the state corresponding to two isolated macroions with $R \longrightarrow \infty$. This SC free energy difference, Eq \ref{eq:path}, finally assumes the form:
 \begin{equation}
 \beta{\cal{F}}=-\ln{[{\cal{Z}}_0]}-\lambda_c\frac{ {\cal{Z}}_1}{{\cal{Z}}_0}={\cal{\tilde{F}}}_0+c_0 {\cal{\tilde{F}}}_1.
 \end{equation} 
Here, in the grand canonical ensemble, the fugacity $\lambda_c$ is identical to polyvalent ion concentration in the bulk $c_0$, and 
\begin{equation}
{\cal{\tilde{F}}}_0 = -\ln{[{\cal{Z}}_0]},
\end{equation}
where ${\cal{Z}}_0$ is defined in Eq. \ref{aeknux}, and extensively analyzed in Ref. \cite{Natasa1,Natasa2}. For the sake of completeness we nevertheless write it down in an explicit form
\begin{eqnarray}
{\cal{\tilde{F}}}_0 &=& -\frac{N[ (pH-pK)\ln{10}]^2}{2+Nl_B\frac{1}{a}[1+\frac{a}{R}e^{-\kappa R}]} + \nonumber\\
&& {\textstyle\frac12} \ln{\left(1+  {N}\frac{l_B}{2 a}+\frac{l_B^2}{(2 a)^2} N^2 [1-\frac{a^2}{R^2}e^{-2\kappa R}]\right)}.\label{eq:ks0}
\end{eqnarray}
On the other hand, ${\cal{\tilde{F}}}_1$, as defined above yields the final expression 
\begin{widetext}
\begin{eqnarray}
{\cal{\tilde{F}}}_1 &=& \int_V\!\! d{\bf r}_0 \Bigg(
\exp{\!\!\Bigg[\!\frac{qN(pH\!-\!pK)\ln{10} \left( u_{DH}({\bf r}_0,{\bf r}_1)\! +\!u_{DH}({\bf r}_0,{\bf r}_2) \right) }{2+Nl_B\frac{1}{a} [1+\frac{a}{R}e^{-\kappa R}]}\!\!\Bigg]} \times \nonumber\\
&&  \exp{\Bigg[\frac{1}{2}q^2 \Big(C_{11}u_{DH}^2({\bf r}_0,{\bf r}_1) + C_{22} u_{DH}^2({\bf r}_0,{\bf r}_2)-} {2C_{12} u_{DH}({\bf r}_0,{\bf r}_1)u_{DH}({\bf r}_0,{\bf r}_2)  \Big) \Bigg]}-1\Bigg),
\label{eq:f1}
\end{eqnarray}
\end{widetext}
with explicitly subtracted free energy value of  two isolated macroions with $R \longrightarrow \infty$.  The structure of this complicated expression is as follows: the first exponent corresponds to the screened DH interactions of the $q$-valent polyvalent ion with both macroions, whose charge is determined by the bulk $pH$ of the solution and is proportional to $pH - pK$, while the second exponent corresponds to the electrostatic self-interaction of the polyvalent ion in the presence of both macroions. Finally the product of the two expressions needs to be integrated over all the possible positions of the polyvalent ion. The constants $C_{11}$ and  $C_{22}$, Eq \ref{eq:cap}, can be interpreted as generalized self-capacitances and mutual capacitance $C_{12}$ of the macroions, originating in the interaction between the three charged particles. At the end, we subtracted the non-interacting  part of two isolated macroions proportional simply to the volume of the system $V$.

In addition, we note that both ${\cal{\tilde{F}}}_0(R)$ as well as ${\cal{\tilde{F}}}_1(R)$ contain parts which are due to {\sl polyion mediated} interaction between the macroions, proportional to $qN (pH -pK)$, as well as {\sl polyion self-interaction} mediated by the macroions and proportional to $q^2$. The division into a "mean interaction" and "fluctuations" is thus not possible due to the fact that our theory is not of a mean-field type that would allow for fluctuations around the mean-field configuration. 

In the case of absent charge regulation, where the system consists of two macroions with fixed charge $Ne_0$, immersed in the same bathing solution with a strongly coupled oppositely charged polyvalent ion, one can repeat the above analysis and obtain the final free energy in the form
\begin{widetext}
\begin{eqnarray}
{\cal{\tilde{F}}}_0+c_0 {\cal{\tilde{F}}}_1= N^2 l_B\left(\frac{1}{a}+\frac{e^{-\kappa R}}{R} \right)-c_0 \int_V\!\! d{\bf r}_0\Bigg( \exp{\Bigg[qN\left( u_{DH}({\bf r}_0,{\bf r}_1) +u_{DH}({\bf r}_0,{\bf r}_2)\right)\!\!\Bigg]}-1\Bigg).
\label{eq:f2}
\end{eqnarray}
\end{widetext}
This is very instructive, since obviously without charge regulation the self-interaction contributions proportional to $q^2$ is absent, and the interaction energy reduces to the macroion-macroion repulsion proportional to the charge squared, $(Ne_0)^2$, and a contribution stemming from the interaction of macro-ions with the polyvalent ion, proportional to the product of both charges, $ q (Ne_0)$.
The above equations represent the final result of the dressed ion theory for the interaction between two identical point-like charge regulated macroions in the presence of small concentrations of a polyvalent salt and they have to be evaluated numerically.

\section{Results and disccussion}

The effective interaction free energy between the charge-regulated macroions is obtained directly from  Eq. \ref{aeknux} after performing the numerical integration over volume in Eq. \ref{eq:f1}.   
We calculate the total interaction free energy, $ \beta{\cal{F}}(R)$, as a function of the separation between the macroions as  
 \begin{equation}
 \beta{\cal{F}}(R)={\cal{\tilde{F}}}_0(R)+c_0 {\cal{\tilde{F}}}_1(R),
 \end{equation}
We study the separation dependence for different values of the parameters, differentiating in particular the case of $pH - pK = 0$, i.e. the point of zero charge (PZC), corresponding to macroions that are on the average uncharged. Inspite of this, the self-energy of the polyvalent ion in this case still contains the non-vanishing electrostatic self-interaction of the polyvalent ion mediated by both charge regulated macroions.

\begin{figure}[t]
\centering{\includegraphics[width=0.45\textwidth]{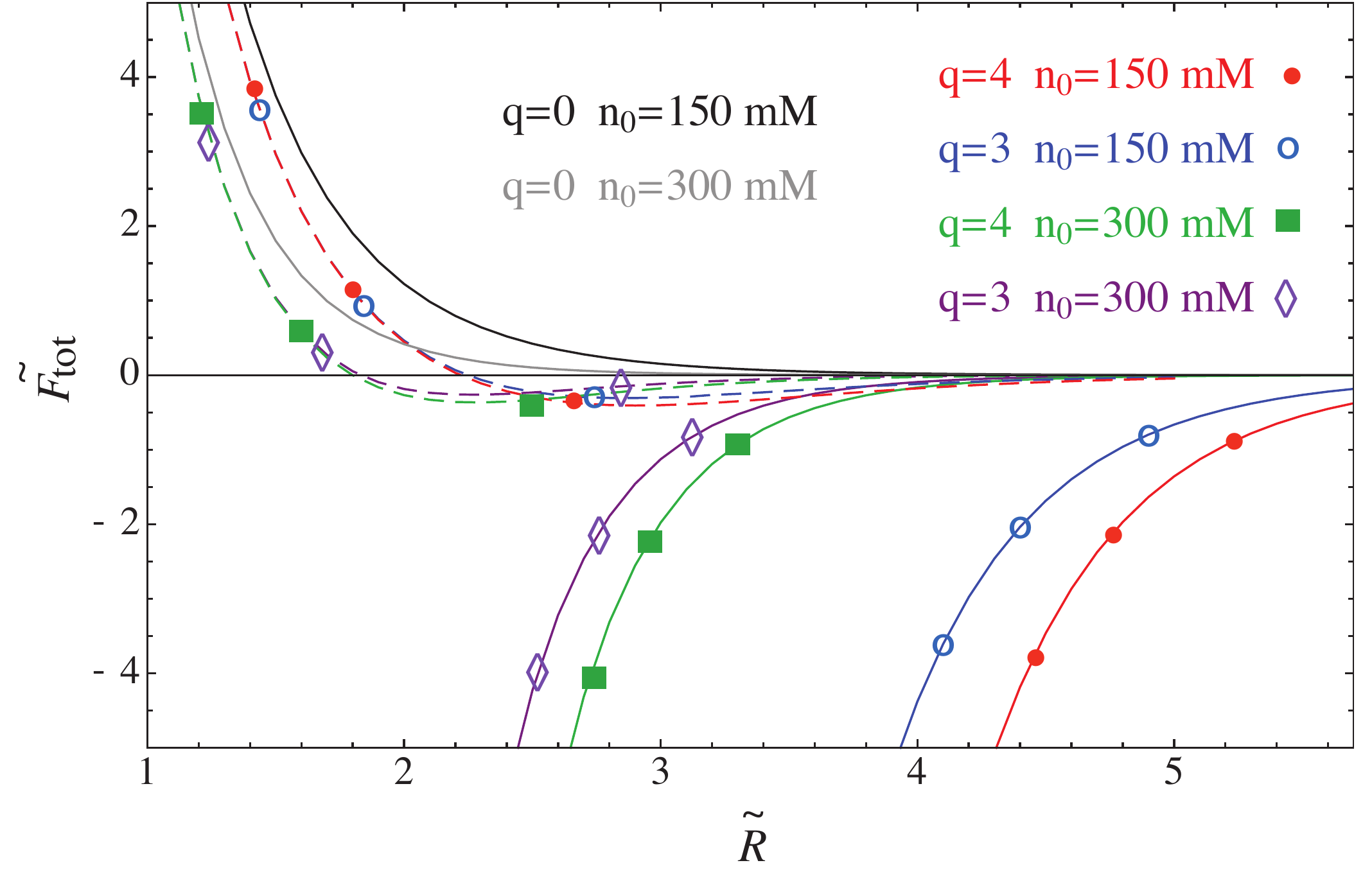}}
\caption{The total interaction free energy for $pH-pK=3$ (solid lines) and $pH-pK=-3$ (dashed lines), at fixed values of parameters as shown in legend.  Macroions of diameter $a=1$ nm, with the number of adsorption sites $N=7$ and salt concentration $c_0=1$ mM. }
\label{fig:fig5}
\end{figure}

\begin{figure}
\centering{\includegraphics[width=0.45\textwidth]{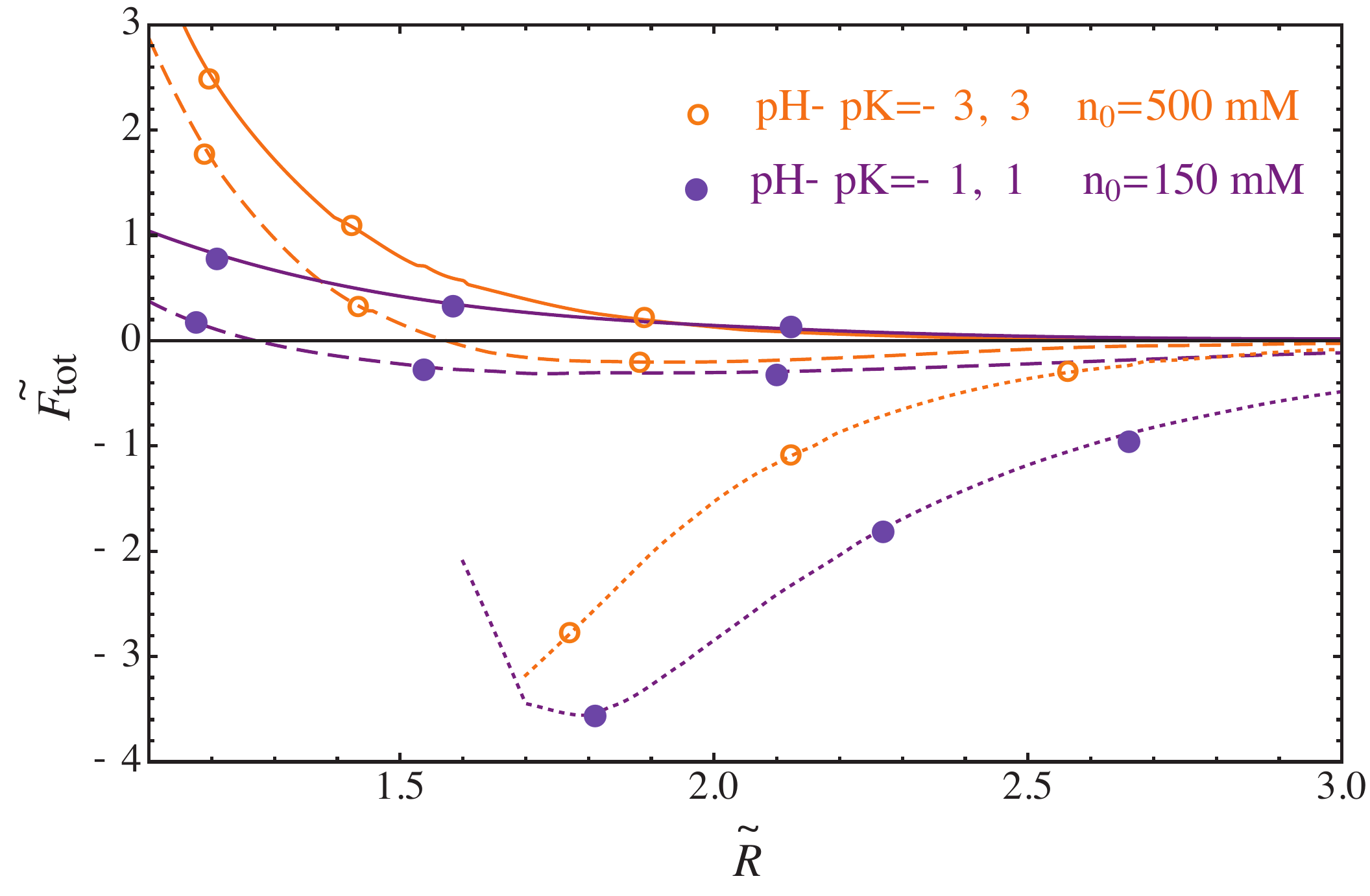}}
\caption{Total interaction free energy between macroions with small $pH - pK$ values at low salt concentration  compared with the total interaction free energy in concentrated salt solutions between macroions with large $pH - pK$. Dotted lines correspond to $q$ being a counter ion, dashed $q$ is coion, while solid lines stand for $q=0$.  Macroions diameter $a=1$ nm, number of adsorption sites $N=7$ and $c_0=1$ mM.}
\label{fig:fig6}
\end{figure}

We first analyze the term ${\cal{\tilde{F}}}_1(R)$ from Eq. \ref{eq:f1}, which corresponds to the interaction free energy mediated by the polyvalent $q$-ion only. Obviously, see Figure \ref{fig:fig2}(a),  this interaction free energy leads to an attractive contribution to the force at PZC, stemming solely from the self-interaction of the polyvalent ion, mediated by the charge regulation of the macroions, whose magnitude depends quadratically on $q$. The screening effect of the monovalent salt is clearly discernible. In summary, the polyvalent self-interaction at PZC yields an attractive interaction that gets stronger and more long-ranged on increase of the valency $q$ of the polyvalent ion and on decrease of the monovalent salt concentration $n_0$. We should note that this PZC polyvalent ion-mediated attraction in the SC dressed ion approach is much stronger then the residual WC (KS) attraction between charge regulated macroions in a monovalent salt solution (black lines) Fig \ref{fig:fig2}(a).

We have not specified yet the sign of the $q$ polyvalent ion. In fact the product $q (pH - pK)$ can have either sign.  In Fig. \ref{fig:fig2} we thus study how the sign of polyvalent ions modifies the polyvalent ion-mediated contribution to the total interaction free energy. For both cases,  $q (pH - pK)$  positive, Fig. \ref{fig:fig2}(b), and for $q (pH - pK)$ negative, Fig. \ref{fig:fig2}(c), the interaction free energy corresponds to attractive polyvalent ion-mediated forces but of vastly different magnitude, being much larger in the former case then in the latter.  In both cases the attraction is again larger in the lower screening regime (less $n_0$, bigger $q$). 

The total interaction free energy between the two titratable macroions, $\beta{\cal{F}}(R)={\cal{\tilde{F}}}_0(R)+c_0 {\cal{\tilde{F}}}_1(R)$, is presented in Figs. \ref{fig:fig5} and \ref{fig:fig6}. Obviously, the interaction force is attractive when $q(pH - pK) \geq 0$, due to the strongly coupled polyvalent ion mediated interaction, and is in general screened by the monovalent salt. Interestingly enough, in this case even the interaction at small separations remains attractive and the bare macroion repulsion is not observed. The reason for this is not the polyion mediated electrostatic attraction but its size: in fact for small separation the polyvalent counterion can not enter the space between the macroions and thus exerts an additional effective osmotically generated attraction between them in general akin to the depletion effect, already noticed in a similar context for net-neutral surfaces at small separations \cite{SCdressed3}. In the opposite case, when $q (pH - pK) < 0$, the repulsion in general prevails, except at large separations where one can detect a small residual attraction, possibly as a consequence of an asymmetrical charge fluctuation due to charge regulation. At smaller separations the bare repulsion between macroions is reduced partly due to the charge regulation effects and partly due to depletion effects. In Fig. \ref{fig:fig6} one can additionally notice how the two cases, one with small $pH - pK$, immersed in a solution of low salt concentration, and the other one with large $pH - pK$, but immersed in concentrated salt solution, have quite similar behavior, indicating that the valency of the polyion and the screening of the monovalent salt somehow act in parallel.

\begin{figure}
\centering{\includegraphics[width=0.46\textwidth]{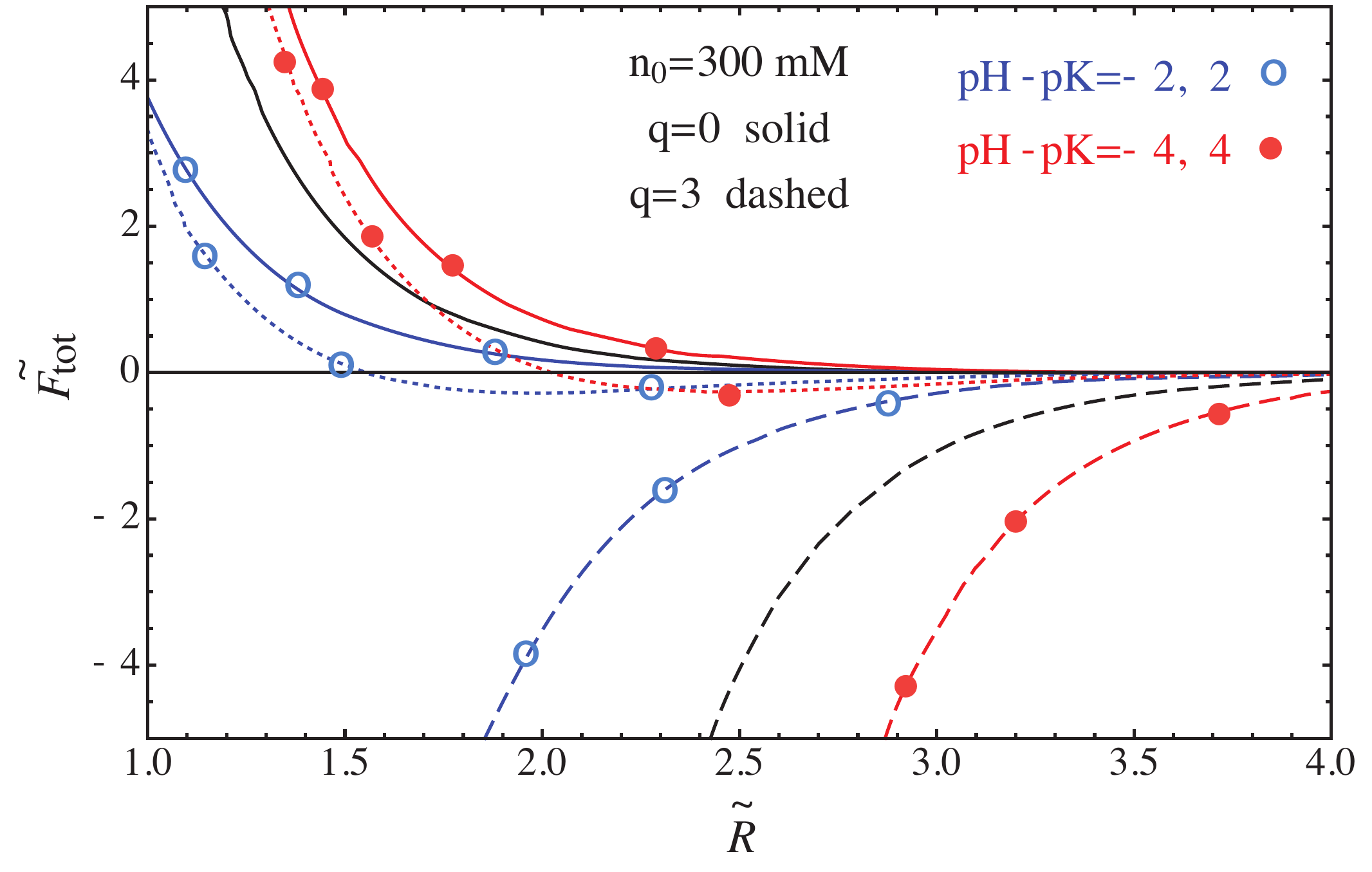}
\caption{The comparison of the total interaction free energy for non-regulated case (black full line) with total regulated interaction energy at $q(pH-pK)<0$ (dotted lines) and $q(pH-pK)>0$ (dashed lines), at fixed values of parameters as shown in legend.  Macroions of diameter $a=1$ nm, with the number of adsorption sites $N=7$ and salt concentration $c_0=1$ mM. }\label{fig:fig7}}
\end{figure}

In  Fig. \ref{fig:fig7}  the total interaction energy  is now compared for the two cases with and without charge regulation, Eq. \ref{eq:f1} and Eq. \ref{eq:f2}, respectively. The charge non-regulated case corresponds to fixed values of the macroion charge equal to $N e_0$. Here, one can notice the important effect of charge regulation through the polyvalent mediated interaction, ruled by the $pH$ value, which determines the overall strength of the charge regulation interaction, that can then appear as either smaller or larger than the one corresponding non-regulated interaction energy. This non-monotonic effect of charge regulation hinges on the two terms in the dressed ion free energy that respond differently to titration of the macroion charges. 

The dressed ion theory obviously predicts an attractive interaction between charge regulated macroions, which can sometimes dominate the overall interaction. This is different from the WC case~\cite{Natasa1,Natasa2}, where the fluctuation attraction, or the KS interaction, is subdominant to the DH repulsion, except close to the PZC, where it indeed becomes dominant. In the SC dressed ion theory the attraction can clearly become dominant either with or without the charge regulation, though it can be stronger in the latter case and remains important for any value of pH. The salt effect acts mostly to quench the correlation polyvalent ion-mediated attraction and diminish its spatial range. 

The attraction between two identical charge regulated macroions, seen in the dressed ion theory, has a different origin from the WC KS interactions, where they are due to thermal monopolar charge fluctuations around the mean-field solution, enabled by the dissociation equilibrium of the surface of the macroion. In the dressed ion theory, the polyvalent ion-mediated attraction could be seen as being due to the {\sl electrostatic bridging interaction} involving the polyvalent ion. 
This should in general not be confused with the s.c. {\sl salt bridging interaction} sometimes invoked even in weakly coupled monovalent salt solutions. 

\section{Conclusion}

The main goal of this research was to present  a theoretical description for the phenomenon of charge regulation as affected by  the presence of polyvalent ions. We formulated a SC dressed ion theory, describing the electrostatic interactions between macroions  undergoing charge regulation processes, in a mixture of monovalent-polyvalent salt solution.  Using the proper description of charge regulation, suitable for treating it in the field theoretical framework, the partition function is derived in the form of a virial expansion valid for small concentration of the polyvalent salt. The first term in such expansion corresponds to the direct interaction between  titratable macroions in a monovalent salt solution, while the first order correction, stems from the interaction of the polyvalent ion with each macroion. The asymmetry in the ionic solution allowed us to decouple the system into the monovalent salt component, addressed on a week coupling level, while the polyvalent ion component was assumed to be strongly coupled with macroions. In both cases, titration of the macroions is treated on the Gaussian approximation level involving an expansion of the exact charge regulation free energy valid in general for highly charged macroions

We have shown that the presence of polyvalent ion brings about a strong attraction between two symmetrically charged macroions. In the case when polyvalent ion acts like a counterion, the attraction is big enough to overcome repulsion between the macroions, while in the opposite case, the repulsion between macroions turns into a small attraction at large separations due to the asymmetric charge fluctuations at macroions surface, induced by the presence of the polyvalent salt.  The polyvalent-ion mediated attraction remains appreciable even at conditions, when macroions reach the point of zero charge.  From the derived expressions for the free energy of interaction, it is clear that the polyvalent ion-mediated attractive contribution stems from the charge-induced charge type of the interaction, since it is proportional to the square of the polyvalent ion charge. Our results show that the polyvalent ion-mediated attraction is significantly stronger then the KS interaction, obtained for the same system described in the WC regime, i.e. without any polyvalent salt. We therefore derived a generalized form of the KS interaction, with the range of validity extended to the regime, where their original KS derivation fails.

By calculating the interaction between point-like charge regulated macroions in the WC and SC approximations, based within the field representation of the partition function, we have opened a new way to analyze the interactions between proteins in ionic solutions. Our approach brings together the charge regulation theory as well as the general  WC and SC dichotomy of the field representation of the partition function of Coulomb fluids. The results seem interesting and we will endeavor to compare them with detailed Monte Carlo simulations in the near future.

\section{Appendix}

\subsection{Fluctuating electrostatic potential propagator}\label{subsec:app1}

The propagator $G_1(\varphi_1,\varphi_2)$, describing how the electrostatic potential propagates from one macroion to another in the presence of a polyvalent ion $q$ at ${\bf r}_0$, is given as:
\begin{widetext}
\begin{eqnarray}
G_1(\varphi_1,\varphi_2)&=& \int{\cal D}[\varphi({\bf r})]\delta(\varphi({\bf r}_1)-\varphi_1)\delta(\varphi({\bf r}_2)-\varphi_2) e^{-\frac{1}{2}\int d{\bf r}d{\bf r} ' \varphi({\bf r})u_{DH}^{-1}({\bf r}, {\bf r} ' )\varphi({\bf r} ' )+i\beta \int_V \rho({\bf r}) \varphi({\bf r}) d{\bf r}}
\end{eqnarray}
\end{widetext}
with $\rho=q\delta ({\bf r}-{\bf r}_0)$. The delta function entering the above expression can be written via a Fourier integral representation  as:
\begin{eqnarray}
\delta(\varphi({\bf r}_i)-\varphi_i)\!&=&\!\int dke^{ik(\varphi({\bf r}_i)-\varphi_i)}\!= \nonumber\\
&& = \!\int dk e^{-ik\varphi_i+ik\int d{\bf r}\rho_i({\bf r})\varphi({\bf r})}\nonumber\\
~
\end{eqnarray}
where $\ \rho_i({\bf r})=\delta({\bf r}-{\bf r}_1)$, $i=1, 2$. One notes that this is an ordinary and not a functional Fourier integral representation, as the propagator is defined for two vertex points in the real space.  Our strategy now will be to first evaluate the functional integral over the fluctuating electrostatic potential field $\varphi({\bf r})$ and then calculate the remaining integral over the auxiliary fields stemming from the Fourier representation of the delta functions.
Therefore it follows that
\begin{widetext}
\begin{eqnarray}
G_1(\varphi_1,\varphi_2) =\! \int \!\!dk_1 e^{-ik_1\varphi_1}\!\!\int \!\!dk_2 e^{-ik_2\varphi_2}\!\!\int\!{\cal D}[\varphi({\bf r})] 
\exp{\Big[\!-\!\frac{1}{2}\int d{\bf r}d{\bf r} ' \varphi({\bf r}) u_{DH}^{-1}({\bf r}, {\bf r} ' )\varphi({\bf r} ' )+} 
{i\int [t({\bf r})+\beta q e_0\delta ({\bf r}-{\bf r}_0)]\varphi({\bf r})d^3{\bf r}\Big]}\nonumber\\
~
\end{eqnarray}
\end{widetext}
with the field $t({\bf r})$ denoting $$\ t({\bf r})=k_1\rho_1({\bf r})+k_2\rho_2({\bf r}).$$The above integral is a general Gaussian functional integral for the fluctuating potential $\varphi({\bf r})$ and can be evaluated explicitly and exactly. The result is then an ordinary Gaussian integral over the variables $k_1$ and $k_2$. 

One has in fact
\begin{eqnarray}
&&\delta(\varphi({\bf r}_1)-\varphi_1)\delta(\varphi({\bf r}_2)-\varphi_2) = \nonumber\\
&& \int dk_1 e^{ik_1(\varphi({\bf r}_1)-\varphi_1)} \int dk_2 e^{ik_2(\varphi({\bf r}_2)-\varphi_2)}
\end{eqnarray}
after which one can derive
\begin{widetext}
\begin{eqnarray}
G_1(\varphi_1,\varphi_2) &=& 
{{\det{u_{DH}^{-1/2}({\bf r},{\bf r}')}}} ~\int dk_1e^{-ik_1\varphi_1}\int dk_2e^{-ik_2\varphi_2} e^{-\frac{1}{2}\int d{\bf r}d{\bf r}'[t({\bf r})+\beta q e_0\delta ({\bf r}-{\bf r}_0)] u_{DH}({\bf r},{\bf r}')[t({\bf r}')+\beta q e_0\delta ({\bf r}'-{\bf r}_0)]}=\nonumber\\
&&= {{\det{u_{DH}^{-1/2}({\bf r},{\bf r}')}}}~\int_{-\infty}^{+\infty}\int_{-\infty}^{+\infty}\!\!\!dk_1dk_2e^{-ik_1\varphi_1-ik_2\varphi_2}\times e^{-\frac{1}{2}k_1^2 u_{DH}({\bf r}_1,{\bf r}_1) -\frac{1}{2}k_2^2 u_{DH}({\bf r}_2,{\bf r}_2) -k_1k_2 u_{DH}({\bf r}_1,{\bf r}_2)}\times \nonumber\\
&& \times e^{-\frac{1}{2}\beta^2 q^2 e_0^2 u_{DH}({\bf r}_0,{\bf r}_0)  -\frac{1}{2}\beta qe_0[2k_1u_{DH}({\bf r}_0,{\bf r}_1) +2k_2 u_{DH}({\bf r}_0,{\bf r}_2)] }.
\end{eqnarray}
\end{widetext}
The fluctuating electrostatics potential propagator has thus been reduced to simple integrals in the variable ${\bf k} = (k_1, k_2)$.

The vacuum fluctuations term, ${{\det{u_{DH}^{-1/2}({\bf r},{\bf r}')}}}$, as well as the polyvalent ion bare self-interaction term $e^{-\frac{1}{2}\beta^2 q^2 e_0^2 u_{DH}({\bf r}_0,{\bf r}_0) }$,  will be neglected since they do not depend on the separation between the point-like macroions and thus make no contribution to the interactions between them. If the macroions had finite dimensions ${{\det{u_{DH}^{-1/2}({\bf r},{\bf r}')}}}$ would describe the thermal Casimir (van der Waals) interactions between them.

If one introduces a 2D wave-vector ${\bf k}$, together with the Einstein summation convention, this integral can be rewritten simply as
\begin{equation}
G_1(\varphi_1,\varphi_2)= \int \!\!\int \!d^2{\bf k} ~e^{- f({\bf k})}
\end{equation}
where we introduced the function $f({\bf k})$ as
\begin{equation}
f({\bf k}) = k_j (i\varphi_j+\beta qe_0 u_{DH}({\bf r}_0,{\bf r}_j))\!+\!{\textstyle\frac{1}{2}} k_j u_{DH}({\bf r}_j,{\bf r}_l) k_l 
\end{equation}
Since this is a Gaussian integral, it can be evaluated explicitly as
\begin{widetext}
\begin{eqnarray}
G_1(\varphi_1,\varphi_2)={{\det{u_{DH}^{-1/2}({\bf r},{\bf r}')}}} \exp{\Big[{\textstyle\frac{1}{2}}\left(i\varphi_i +\beta qe_0 u_{DH}({\bf r}_0,{\bf r}_i) \right) {u_{DH}^{-1}}({\bf r}_i,{\bf r}_j) \left( i\varphi_j +\beta qe_0 u_{DH}({\bf r}_0,{\bf r}_j)\right)\Big]}\end{eqnarray}.
\end{widetext}
The above expressions typically involve the Coulomb or the DH self-interaction $u_{DH}({\bf r},{\bf r})$, or indeed its inverse. This quantity is not unambiguously defined because the field representation does not describe the sizes of the charges in a consistent description. However, one usually assumes that the finite size can be approximately included as an ultraviolet cutoff in the Fourier space, or indeed by assuming that one has the Coulomb self-energy $u_{DH}({\bf r},{\bf r}) \sim 1/ 4\pi \varepsilon\varepsilon_0 a$, where $a$ is the radius of the charge; to be consistent one needs to take $\kappa a \longrightarrow 0$ in the DH expression, which gives its bare Coulomb limit.

\section{Acknowledgements}

N. A. acknowledges the financial support by the Slovenian Research Agency under the young researcher grant. R.P. acknowledges the financial support by the Slovenian Research Agency under the grant P1-0055. The authors would also like to thank Profs. David Andelman and Michal Borkovec as well as Drs. Tomer Markovich and Gregor Trefalt for illuminating discussions on the subject of charge regulation and electrostatic interactions between charge-regulated macroions.

\section{REFFERENCES:}

\end{document}